\documentclass[10pt,twoside,twocolumn,english,aps,manuscript,aps,preprint,showpacs,superscriptaddress,showkeys]{revtex4}
\usepackage{lmodern}
\usepackage[T1]{fontenc}
\usepackage[latin9]{inputenc}
\usepackage{textcomp}
\usepackage{amstext}
\usepackage{graphicx}

\makeatletter

\newcommand{\lyxmathsym}[1]{\ifmmode\begingroup\def\b@ld{bold}
  \text{\ifx\math@version\b@ld\bfseries\fi#1}\endgroup\else#1\fi}

\providecommand{\tabularnewline}{\\}

\@ifundefined{textcolor}{}
{%
 \definecolor{BLACK}{gray}{0}
 \definecolor{WHITE}{gray}{1}
 \definecolor{RED}{rgb}{1,0,0}
 \definecolor{GREEN}{rgb}{0,1,0}
 \definecolor{BLUE}{rgb}{0,0,1}
 \definecolor{CYAN}{cmyk}{1,0,0,0}
 \definecolor{MAGENTA}{cmyk}{0,1,0,0}
 \definecolor{YELLOW}{cmyk}{0,0,1,0}
}

\makeatother

\usepackage{babel}
\begin{document}

\title{Dual role of an ac driving force and the underlying two distinct
order-disorder transitions in the vortex phase diagram of Ca$_{3}$Ir$_{4}$Sn$_{13}$}

\author{Santosh Kumar}

\email{santoshkumar@phy.iitb.ac.in}

\address{Department of Physics, Indian Institute of Technology Bombay, Mumbai
400076, India}

\author{Ravi P. Singh}

\altaffiliation{Department of Physics, Indian Institute of Science Education and Research, Bhopal 462066, India.}

\address{Department of Condensed Matter Physics and Materials Science, Tata
Institute of Fundamental Research, Mumbai 400005, India.}

\author{A. Thamizhavel}

\address{Department of Condensed Matter Physics and Materials Science, Tata
Institute of Fundamental Research, Mumbai 400005, India.}

\author{C. V. Tomy}

\address{Department of Physics, Indian Institute of Technology Bombay, Mumbai
400076, India}

\author{A. K. Grover}

\address{Department of Condensed Matter Physics and Materials Science, Tata
Institute of Fundamental Research, Mumbai 400005, India.}

\address{Department of Physics, Panjab University, Chandigarh 160014, India.}
\begin{abstract}
We present distinct demarcation of the Bragg glass (BG) to multi-domain
vortex glass (VG) transition line and the eventual amorphization of
the VG phase in a weakly pinned single crystal of the superconducting
compound Ca$_{3}$Ir$_{4}$Sn$_{13}$ on the basis of comprehension
of the different yields about the second magnetization peak (SMP)
anomaly in the dc magnetization and the corresponding anomalous feature
in the ac susceptibility measurements. The shaking by a small ac magnetic
field, inevitably present in the ac susceptibility measurements, is
seen to result in contrasting responses in two different portions
of the field-temperature (H, T) phase space of the multi-domain VG.
In one of the portions, embracing the BG to VG transition across the
onset of the SMP anomaly, the ac drive is surprisingly seen to assist
the transformation of the well ordered BG phase to a lesser ordered
VG phase. The BG phase exists as a superheated state over a small
portion of the VG space and this attests to the first order nature
of the BG to VG transition.
\end{abstract}



\maketitle

\section{Introduction}

The advent of high $T_{c}$ superconductivity \cite{key-1} had made
it enlightening and relevant to locate different phase boundaries
in the $H$--$T$ phase space \cite{key-2,key-3,key-4,key-5,key-6,key-8,key-7,key-9},
depicting the various possible phase transformations for the vortex
matter in a type-II superconductor. A widely investigated anomaly
elucidating the features of the vortex phase diagram of a high $T_{c}$
cuprate superconductor, YBa$_{2}$Cu$_{3}$O$_{7}$ (YBCO) \cite{key-9,key-10,key-11,key-12,key-13,key-15,key-14,key-16},
has been the peak effect (PE) phenomenon \cite{key-2}, which amounts
to a sudden increase in the otherwise monotonically decreasing critical
current density ($j_{c}$) just before reaching the upper critical
field line, $H_{c2}(T)$. As per the Larkin-Ovchinnikov \cite{key-17}
description, $j_{c}$ is related inversely to the volume ($V_{c}$)
($j_{c}\propto1/\lyxmathsym{\textsurd}V_{c}$) of the domain within
which the displacements of the individual flux lines remain well correlated.
Hence, an anomalous increase in $j_{c}(H,T)$ (as in PE phenomenon)
signifies a shrinkage in $V_{c}$, which in turn, amounts to an order-disorder
transformation of the vortex matter. Somewhat analogous to the PE,
there is another anomalous feature occurring deep inside the mixed
state (well below $H_{c2}$), called the second magnetization peak
(SMP) \cite{key-15,key-18,key-19,key-20,key-21,key-22,key-23,key-9,key-16,key-24}
which also imprints as a non-monotonic behavior of $j_{c}(H)$. The
SMP feature has been well studied in high $T_{c}$ cuprate superconductors,
YBa$_{2}$Cu$_{3}$O$_{7}$ (YBCO) \cite{key-9}, Bi$_{2}$Sr$_{2}$CaCu$_{2}$O$_{8}$
(BSCCO) \cite{key-20,key-21} and another oxide superconductor, (Ba,K)BiO
system\cite{key-22,key-23,key-24}. 

As an important consequence of the above mentioned studies of the
vortex phase transformations in the high $T_{c}$ superconductors,
the PE/SMP behavior and the vortex phase diagrams have been revisited
various times in several conventional superconductors, such as, Nb
\cite{key-25,key-26,key-27}, 2H-NbSe$_{2}$ \cite{key-2,key-3,key-28,key-29,key-30,key-31,key-32,key-7,key-33},
YNi$_{2}$B$_{2}$C \cite{key-34,key-35}, LuNi$_{2}$B$_{2}$C \cite{key-34}
Ca$_{3}$Rh$_{4}$Sn$_{13}$ \cite{key-36,key-37,key-38,key-16,key-39},
Yb$_{3}$Rh$_{4}$Sn$_{13}$ \cite{key-40,key-41} etc., The order-disorder
transitions (\textit{a la} PE and SMP) reported in Ca$_{3}$Rh$_{4}$Sn$_{13}$
(CaRhSn) by some of the present authors \cite{key-16} bore a marked
resemblance with characteristic results in some single crystals of
the high $T_{c}$ superconductor YBCO. In addition, a recent report
\cite{key-39} of an exemplification of the notion of inverse melting
of the vortex lattice in a crystal of CaRhSn echoes well with the
similar characteristic in the high $T_{c}$ superconductors YBCO \cite{key-42}
and BSCCO \cite{key-43}. The above mentioned investigations in the
low $T_{c}$ superconductors have thus lead to the revelation of many
instructive aspects of the PE and the SMP anomalies. For example,
the results of small angle neutron scattering (SANS) together with
ac susceptibility measurements performed concurrently in a single
crystal of Nb provided convincing evidence of a quasi-first-order
transition, from flux line lattice to amorphous vortex solid, across
the onset of the PE anomaly \cite{key-26}. The SMP on the other hand,
is widely accepted \cite{key-18,key-9,key-19} as a disorder or pinning-induced
transition amounting to a well ordered Bragg glass (BG) state transforming
to a \textit{multi-domain} vortex glass (VG) phase \cite{key-4,key-20,key-44,key-5}.
The first-order nature of BG to VG transition, however, awaits convincing
elucidation. In spite of advances in understanding and continuous
updating of the vortex phase diagrams of a variety of type-II superconductors
\cite{key-32,key-33,key-35,key-45}, several issues still remained
to be fully comprehended. One of these is the delineation between
the SMP and PE anomalies when they juxtapose, and the underlying physics
represented by them. 

Motivated by the aforesaid interesting features reported for a variety
of superconductors, we have now investigated another low $T_{c}$
($\sim7.10$\,K) \cite{key-46,key-47,key-48} superconductor Ca$_{3}$Ir$_{4}$Sn$_{13}$,
which belongs to the family of ternary stannides \cite{key-49} having
cubic structure. This compound has attracted particular attention
in recent years. For example, there have been evidences for nodeless
superconductivity \cite{key-48} and a coexistence of superconductivity
and ferromagnetic spin fluctuations \cite{key-47} (attributed to
Ir 4d-band) in Ca$_{3}$Ir$_{4}$Sn$_{13}$. However, to the best
of our knowledge, studies related to vortex phase transformations
in its superconducting mixed state have not been reported in this
compound. In this work, we present a comparison of different outcomes
of the ac and dc magnetization measurements performed in a weakly
pinned (ratio of depinning and depairing current densities is $\sim10^{-5}$)
single crystal of Ca$_{3}$Ir$_{4}$Sn$_{13}$. The magnetization
data reveal a very broad composite anomalous variation in $j_{c}(H,T)$
which amounts to an order-disorder transformation in the vortex matter,
however, the specific phase boundaries, pertaining to the order-disorder
transformation, obtained from the ac and the dc measurement techniques
are found to be significantly different. We can identify a certain
region of the phase space where the ac field unexpectedly assists
the transformation of the quasi-ordered BG into the partially disordered
VG phase. In a different region of the $H$--$T$ phase space, the
ac drive is seen to enhance the quality of spatial order of the underlying
vortex matter, which inter alia is indicative of another order-disorder
transition occurring in that portion of phase space, which gets exposed
under the influence of an ac field. The two contrasting roles of the
ac drive in different portions are attributed to the juxtaposition
of the SMP anomaly (pinning induced) and the peak effect (PE) phenomenon
(collapse of elasticity induced amorphization of the vortex matter).
The unusual outcome of the imposition of an ac drive near the onset
field of SMP anomaly in the present report also corroborates a very
recent finding \cite{key-50} in a different class of superconductor,
(i.e., in Ba$_{0.5}$K$_{0.5}$Fe$_{2}$As$_{2}$), wherein a large
ac field ($\sim12$ Oe) lowers the onset field value of the anomalous
change in $j_{c}$ at a given temperature, thereby elucidating the
role of ac driving force in facilitating the spatial disordering of
the vortex matter in a portion of the phase-space where the ordered
state is metastable superheated state.

\section{Experimental details}

The single crystals of Ca$_{3}$Ir$_{4}$Sn$_{13}$ were grown by
tin flux method \cite{key-46}. The crystal chosen for the present
study is platelet shaped, with a planar area of $4.7$\,mm$^{2}$
and with thickness of $0.64$\,mm, and mass of $26.2$\,mg. The
ac and dc magnetization measurements were performed in a Superconducting
Quantum Interference Device\,-\,Vibrating Sample Magnetometer (SQUID-VSM,
Quantum Design Inc., USA). The amplitude of the vibration in SQUID-VSM
was kept small for all the dc magnetization measurements so as to
avoid artifact that could arise from possible magnetic field inhomogeneity
of a superconducting magnet along the length of sample movement. The
ac magnetization measurements were performed by superimposing an oscillating
magnetic field on various dc magnetic fields. The peak amplitude and
the frequency of the driving ac field were chosen as $3.5$\,Oe and
$211$\,Hz, respectively.

\begin{figure}[!t]
\begin{centering}
\includegraphics[scale=0.48]{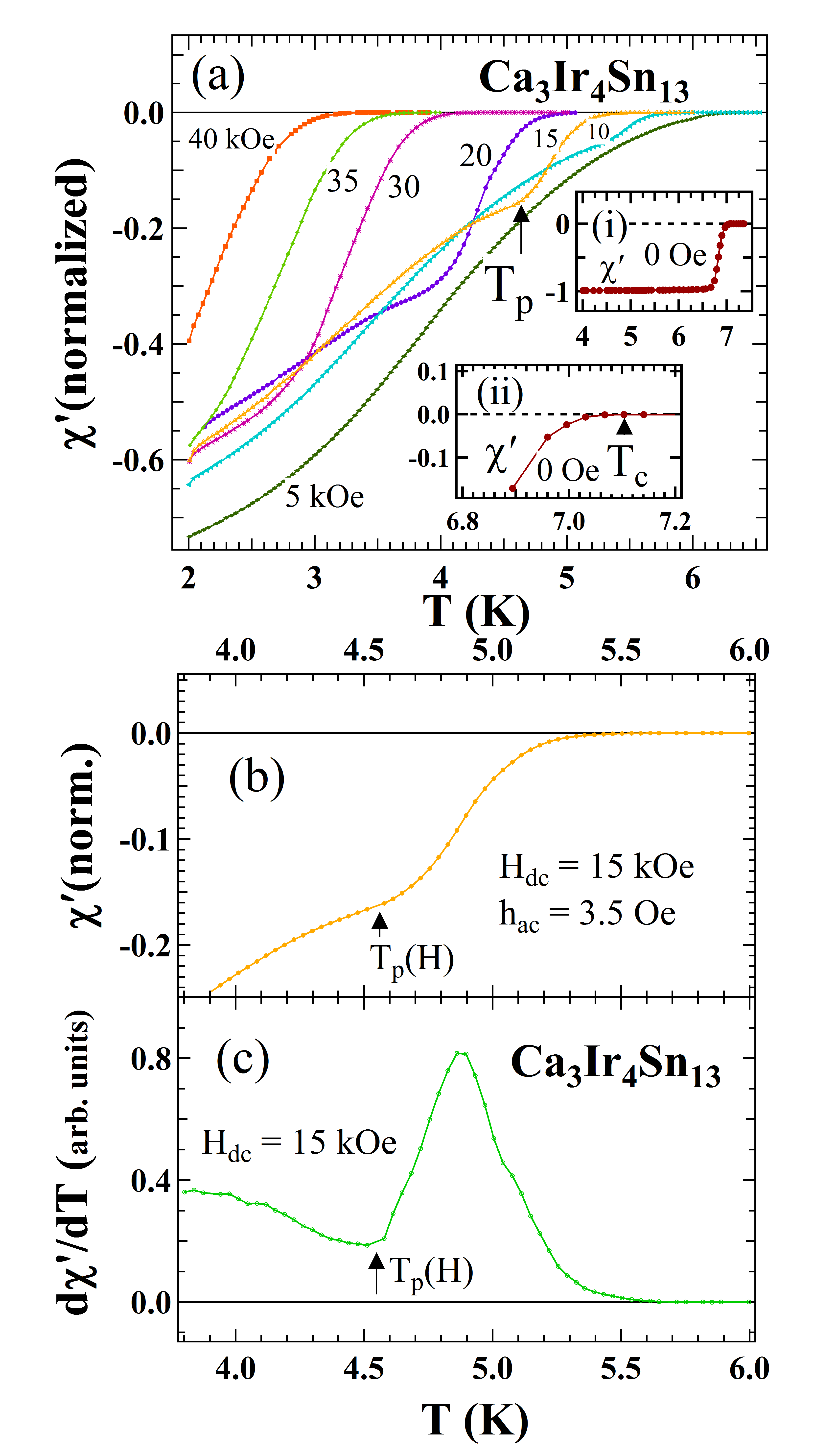}
\par\end{centering}

\caption{{\footnotesize (Color online) (a) Real part of the ac susceptibility,
$\chi^{\prime}(T)$ (in normalized units) plotted against temperature
at various dc magnetic fields (applied parallel to the plane of platelet
shaped crystal and as indicated). Inset (i) shows the saturated $\chi^{\prime}(T)$
response for $H_{dc}=0$, and the inset (ii) displays its magnified
portion near $T_{c}$. (b) Portion of $\chi^{\prime}(T)$ curve for
$H_{dc}=15$\,kOe showing an anomalous modulation prior to the transition
to the normal state. (c) Portion of derivative plot of $\chi^{\prime}(T)$
(with respect to $T$) for $H_{dc}=15$\,kOe locating the temperature
($T_{p}(H)$) corresponding to a position of a maximum in $j_{c}(T)$. }}
\end{figure}

\section{Results}

\subsection{Ac susceptibility: Identification of anomalous variation in critical
current density.}

\subsubsection{Isofield ac susceptibility measurements}

Figure~1(a) shows the temperature dependences of the real part of
the isofield ac susceptibility ($\chi^{\prime}(T)$) in dc fields
(applied parallel to plane of the platelet shaped sample), as indicated.
The $\chi^{\prime}(T)$ plots have been normalized with respect to
the saturated value of $\chi^{\prime}$ recorded at $T=2$\,K for
$H_{dc}=0$. The $\chi^{\prime}(T)$ curve for $H_{dc}=0$ shows the
saturated behavior at low temperatures (see the inset (i) of Fig.~1(a)),
revealing the characteristic full magnetic shielding effect, typical
of a superconductor. The superconducting transition temperature, $T_{c}(H_{dc}=0)\approx7.1$\,K,
obtained via the onset of diamagnetic response (see expanded plot
in the inset (ii) of Fig.~1(a)) from this measurement agrees with
the $T_{c}$ reported for this compound in Ref. \cite{key-47}. The
$\chi^{\prime}(T)$ response for $H_{dc}=5$\,kOe in the main panel
of Fig.~1(a) shows a much broader transition to the normal state
depicting the monotonically decreasing value of $\chi^{\prime}(T)$
as the temperature is increased. Since $\chi^{\prime}$ is related
to the critical current density, $j_{c}(H,T)$ ($(\chi^{\prime}\sim-\beta j_{c}/h_{ac})$
\cite{key-10}, where $h_{ac}$ is the amplitude of the ac drive and
$\beta$ depends on the size and geometry of the sample), the fall
in |$\chi^{\prime}(T)$| reflects a decrease in $j_{c}(H,T)$ with
increasing temperature. However, the $\chi^{\prime}(T)$ response
for $H_{dc}=10$\,kOe shows an unusual deviation from the smooth
monotonic decrease prior to the superconducting to normal transition,
which in turn indicates a possible non-monotonic behavior of $j_{c}$.
This modulation in $\chi^{\prime}(T)$ becomes more pronounced for
$H_{dc}=15$\,kOe (marked by arrow) and for $H_{dc}=20$\,kOe, but
the same characteristic could not be identified clearly at higher
fields ($H_{dc}>25$\,kOe). In order to bring out the interesting
characteristics of this modulation, we now examine in Fig.~1(b),
the behavior of $\chi^{\prime}(T)$ near the region, where this modulation
is observed for $H_{dc}=15$\,kOe. This feature can be more clearly
discerned in an expanded plot of the derivative of $\chi^{\prime}(T)$
with respect to $T$ (see Fig.~1(c)). The point of inflection in
$\chi^{\prime}(T)$, where the slope value is minimum, identifies
the maximum in $j_{c}$ as a function of temperature at a constant
field. This temperature is marked as $T_{p}(H)$ ($\sim4.56$\,K).
After $T_{p}(H)$, the slope $d\chi\prime/dT$ increases abruptly,
reaching a maximum value and then falling to a value close to zero.

\subsubsection{Isothermal ac susceptibility measurements}

\begin{figure}[!t]
\begin{centering}
\includegraphics[scale=0.45]{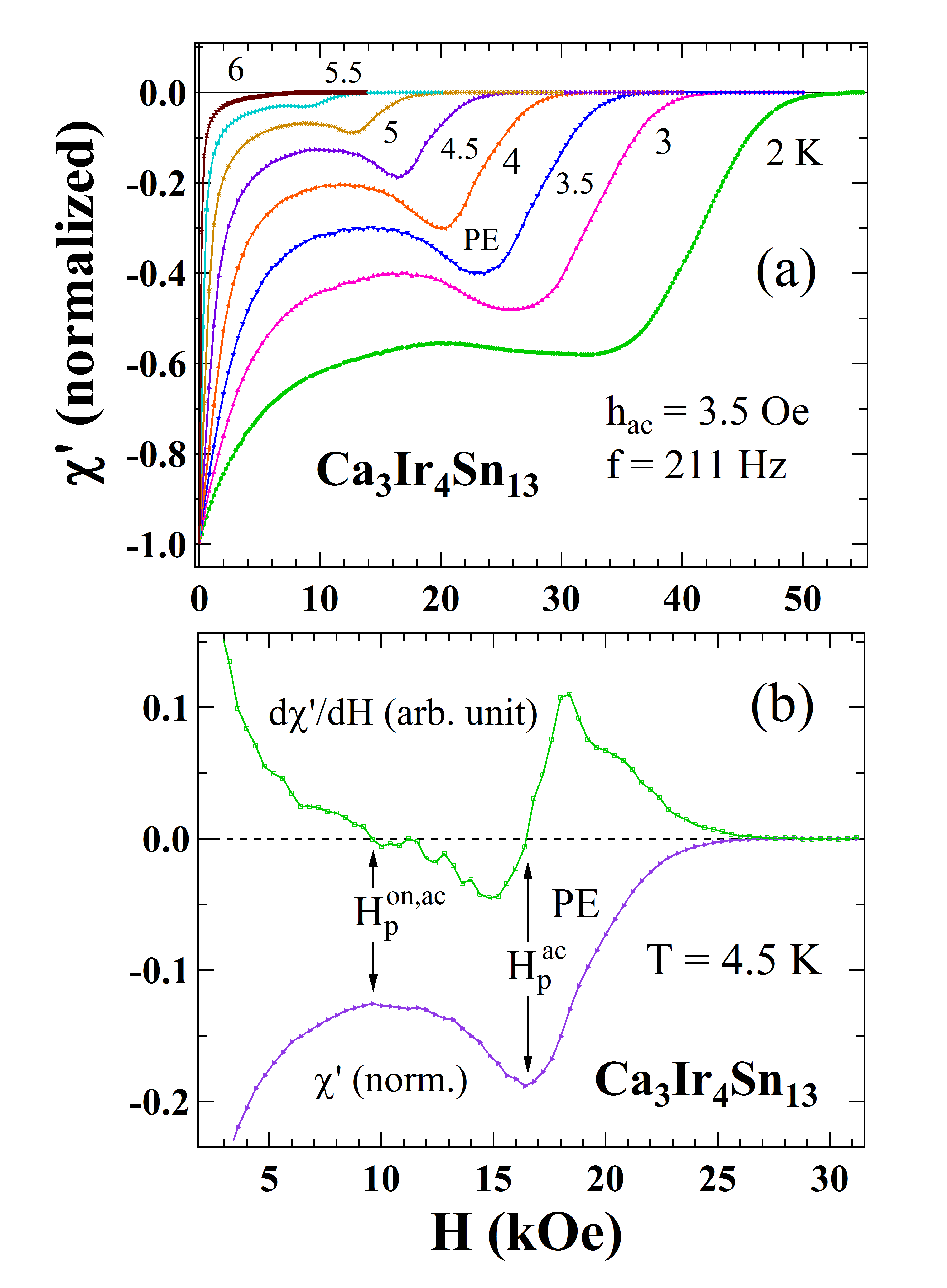}
\par\end{centering}

\caption{{\footnotesize (Color online) (a) Isothermal ac susceptibility ($\chi^{\prime}(H)$)
plotted against the dc magnetic field at various temperatures, as
indicated. PE phenomenon could be observed for $T<5.5$\,K. (b) Portions
of $\chi^{\prime}(H)$ (in normalized units) and $d\chi^{\prime}/dH$
(in arbitrary units) plots at $T=4.5$\,K. The first zero crossing
in $d\chi^{\prime}/dH$ indicates the onset ($H_{p}^{on,ac}$) of
the PE, whereas the second one marks the peak field ($H_{p}^{ac}$)
of the PE. }}
\end{figure}
To further understand the details of the modulation in $j_{c}(H,T)$
reflected in the isofield ac susceptibility measurements, we recorded
isothermal ac susceptibility, $\chi^{\prime}(H)$, data as well, where
the above discussed anomalous trend could be observed more clearly
over a wide range of temperatures ($T\leq5.5$\,K), as shown in Fig.~2(a)
($\chi^{\prime}(H)$ values are normalized with respect to their saturated
value at $H=0$). The specimen was cooled in a nominal zero field
down to a desired temperature and then the $\chi^{\prime}(H)$ data
were recorded while increasing the dc field. The $\chi^{\prime}(H)$
response at each temperature initially decreases as the magnetic field
is enhanced from zero field value, revealing the usual fall in $j_{c}(H)$
with increasing field at a constant temperature. However, an anomalous
dip in $\chi^{\prime}(H)$ response can be seen at higher field values.
As the temperature is increased, the onset field values of this dip
gradually shift towards lower field values (cf. Fig. 2(a)), reaching
closer to the respective critical field values of superconducting-normal
transition, which, in turn, suggests that the observed anomaly is
akin to the well documented peak effect (PE) phenomenon \cite{key-2,key-3,key-7,key-8,key-9,key-10,key-11,key-12,key-13,key-14,key-15,key-25,key-26,key-27,key-28,key-29,key-30,key-31,key-32,key-33,key-34,key-35,key-36,key-37,key-16,key-38}.
In Fig.~2(b), we show an expanded portion of $\chi^{\prime}(H)$
response at $T=4.5$\,K, where the onset field ($H_{p}^{on,ac}$)
of the peak effect like feature and the peak field ($H_{p}^{ac}$)
of the PE are indicated by arrows. We further clarify the PE feature
by plotting the derivative of $\chi^{\prime}$ with respect to $H$
at $T=4.5$\,K in the same panel. The derivative curve ($d\chi^{\prime}/dH$)
in Fig. 2(b) may appear on first examination to be analogous to the
derivative plot $d\chi^{\prime}/dT$ at $H_{dc}=15$\,kOe (cf. Fig.~1(c)).
However, it is to be noted that the derivative of $\chi^{\prime}(H)$
with respect to $H$ in Fig. 2(b) has two zero crossings; the first
one is taken to imprint the onset of peak effect like anomaly ($H_{p}^{on,ac}$)
and the second one identifies the peak position of the PE ($H_{p}^{ac}$).
\begin{figure}[!t]
\begin{centering}
\includegraphics[scale=0.45]{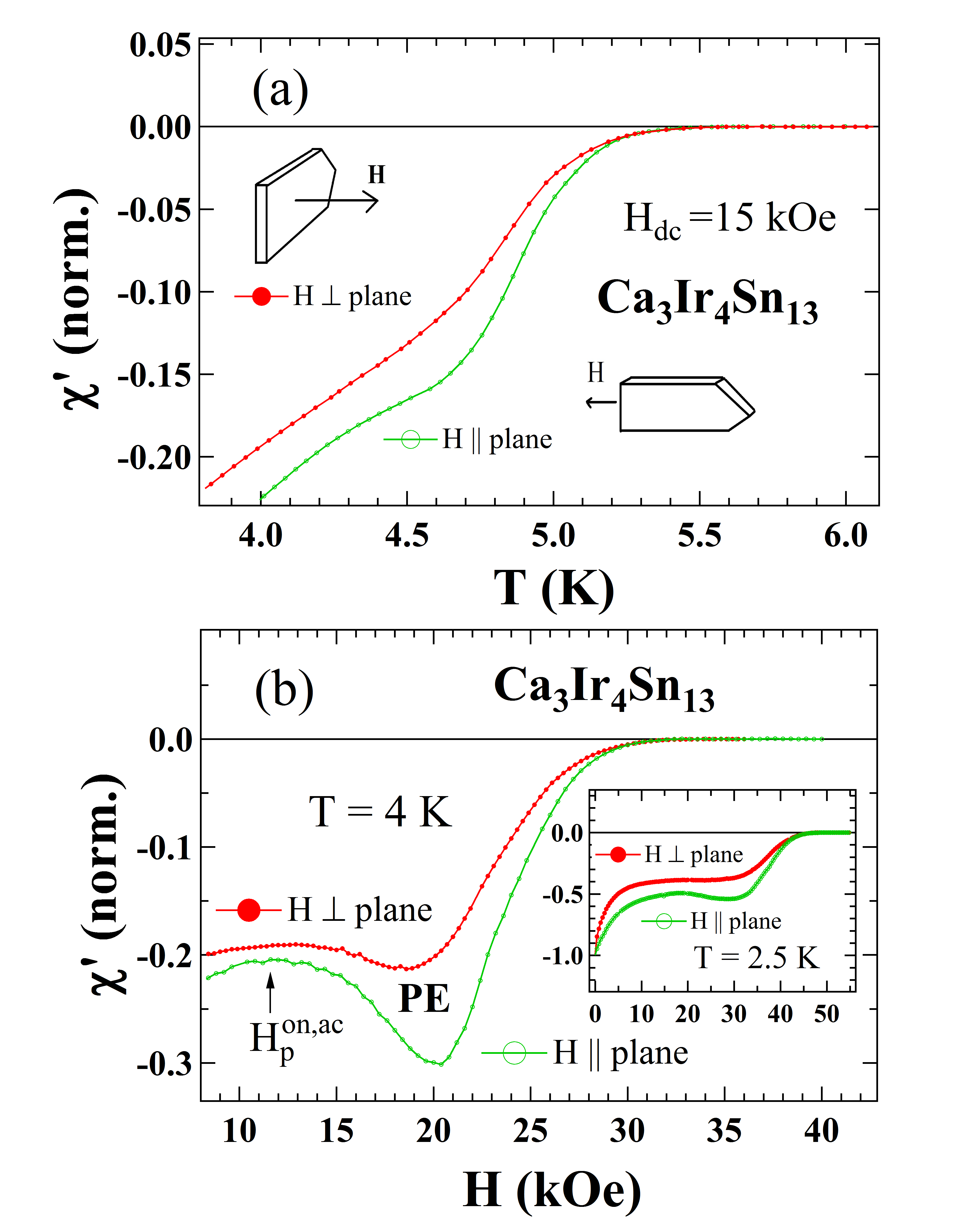}
\par\end{centering}

\caption{{\footnotesize (Color online) (a) Temperature dependences of the ac
susceptibility ($\chi^{\prime}(T)$) for dc field $=15$\,kOe, applied
perpendicular as well as parallel to the plane of the platelet shaped
crystal. The two orientations of the crystal (relative to field) are
drawn in this panel. (b) Isothermal $\chi^{\prime}(H)$ response (normalized)
for these two orientations of the crystal at $T=4$\,K (main panel)
and $T=2.5$\,K (inset panel).}}
\end{figure}

\subsubsection{Sample geometry and anomalous variation in current density}

The sample geometry and demagnetization factor can affect the state
of the spatial order of vortex matter in a given sample due to edge
effects \cite{key-51,key-52}, which in turn could influence the triggering
of the anomalous variation 
\begin{figure}[!b]
\begin{centering}
\includegraphics[scale=0.48]{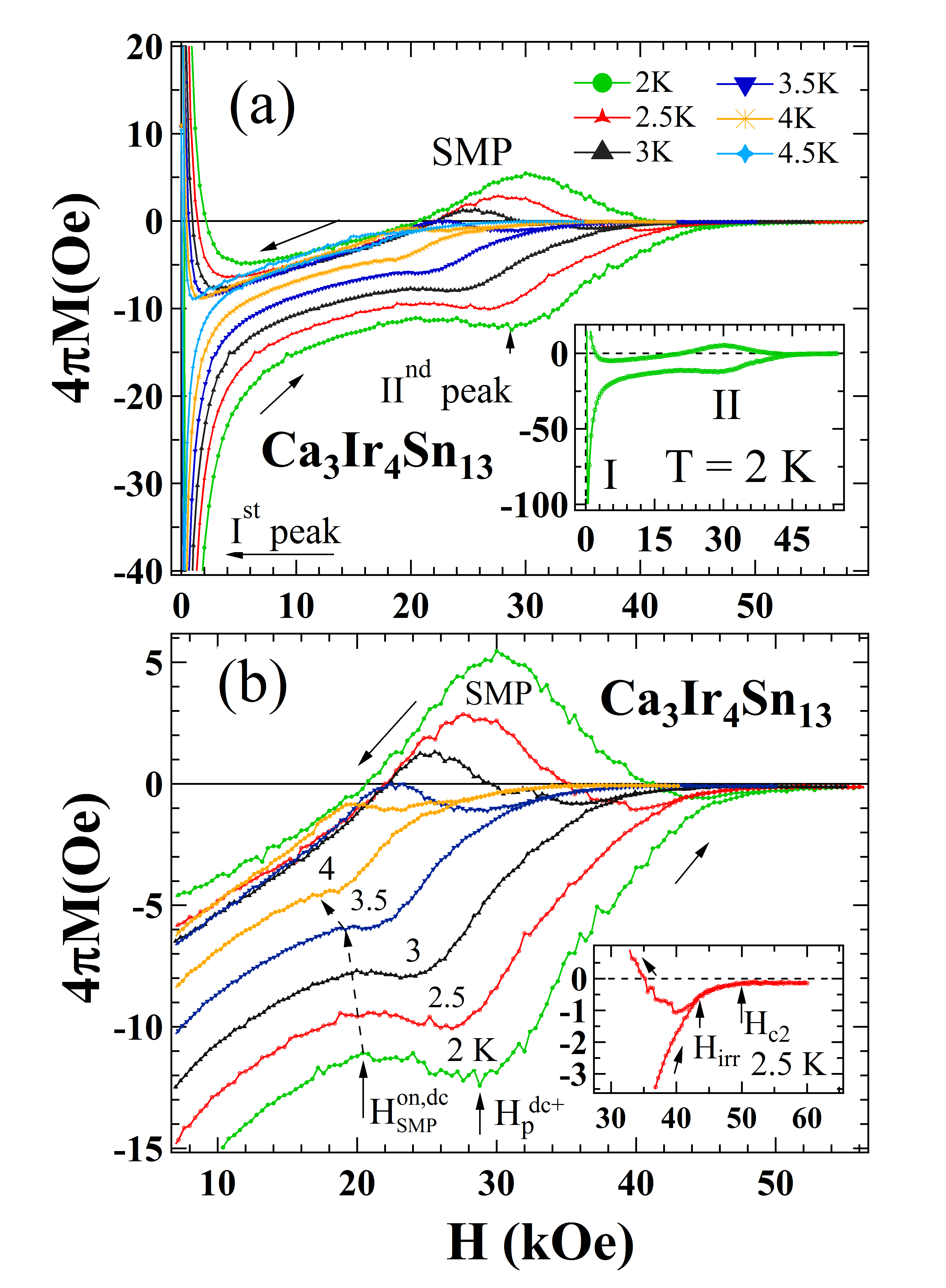}
\par\end{centering}

\caption{{\footnotesize (Color online) (a) Isothermal $M$--$H$ scans (first
two quadrants) recorded at different temperatures, as indicated. Inset
shows one typical $M(H)$ curve recorded at $T=2$\,K with the two
peaks occurring in $M(H)$ curve marked as I and II peak. (b) Expanded
portions of the $M(H)$ loops near an anomalous enhancement in hysteresis
width ($\Delta M=M(H\uparrow)-M(H\downarrow)$) obtained at different
temperatures. The irreversibility field ($H_{irr}$) and the upper
critical field ($H_{c2}$) have been located in the $M(H)$ curve
at $T=2.5$\,K (see inset).}}
\end{figure}
in $j_{c}$ in isofield/isothermal scans. This motivated us to measure
the ac susceptibility with field applied normal to the plane of the
Ca$_{3}$Ir$_{4}$Sn$_{13}$ sample. We show in Fig.~3(a), a comparison
of $\chi^{\prime}(T)$ in the two different orientations: $H$ parallel
and perpendicular to the plane of the sample. It is apparent that
a deviation from the monotonically decreasing $\chi^{\prime}(T)$
is present across nearly identical temperature intervals in both the
orientations. The anomaly is also present in isothermal $\chi^{\prime}(H)$
responses across similar field intervals for both the orientations;
typical curves are shown for $T=4$\,K (Fig.~3(b)) and $T=2.5$\,K
(an inset in Fig.~3(b)). Thus, we may conclude that the above stated
features observed in the ac susceptibility are characteristic of the
pinned vortex matter in the bulk of the Ca$_{3}$Ir$_{4}$Sn$_{13}$
crystal. The anomalous peak feature is somewhat less prominent for
$H$ perpendicular to the plane of the crystal, presumably due to
the pinned vortex state being less spatially ordered prior to the
PE in this orientation. The vortex phase boundaries determined from
the characteristic features of the anomalous variation in $j_{c}$
are however found to be not significantly dependent on the demagnetization
factor and the sample geometry.

\subsection{Identification of the second magnetization peak (SMP): dc magnetization
hysteresis loops.}

\subsubsection{Isothermal M-H scans}

Figure~4(a) displays the first two quadrants of isothermal magnetization
data, i.e., $M$--$H$ loops recorded at various temperatures, for
field applied parallel to the plane of the platelet sample. The sample
was first cooled in nominal zero field down to a desired temperature
and the magnetization was recorded while ramping the field $@$ $100$\,Oe/s
well beyond the upper critical field, and then back to zero. Figure~4(b)
presents the portions of $M$--$H$ loops in the vicinity of an anomalous
enhancement seen in the hysteresis width ($\Delta M(H)=M(H\uparrow)-M(H\downarrow)$).
As per a prescription of the Bean\textquoteright{}s critical state
model \cite{key-53}, the hysteresis width ($\Delta M(H)$) of the
$M(H)$ curve is roughly proportional to $j_{c}(H)$ \cite{key-54}.
Therefore, the second magnetization peak feature \cite{key-15,key-18,key-19,key-20,key-21,key-22,key-23,key-9,key-16,key-24}
observed in the $M(H)$ curves (Fig.~4) reflects an anomalous modulation
in $j_{c}(H)$, which, in turn, could be indicative of an order-disorder
transition in the vortex matter. The $M(H)$ data show that the onset
position (marked as $H_{SMP}^{on,dc}$ in Fig.~4(b)) of this anomaly
is nearly temperature-independent in the range $2$\,K\,$\leq$\,$T$\,$\leq$\,$3.5$\,K
(shown by a dashed arrow line in Fig.~4(b)). The onset field ($H_{SMP}^{on,dc}$)
of the characteristic anomalous feature is located far away from the
threshold field value ($H_{c2}$) of the superconducting-normal transition
for $T<4$\,K, which is in contrast to the quintessential PE phenomenon
typically located at the edge of the $H_{c2}$ in very weakly pinned
superconductors. The anomalous variation in $j_{c}(H)$ located away
from $H_{c2}$ (see inset in Fig. 4(b) for location of $H_{c2}$ at
$2.5$\,K) is termed as the SMP anomaly. The inset in Fig. 4(b) also
shows that the forward ($M(H\uparrow)$ and reverse ($M(H\downarrow)$)
magnetization curves merge at a field identified as the irreversibility
field, $H_{irr}$.

\subsubsection{Magnetization response above the irreversibility field/temperature}

Figure~5(a) illustrates a portion of the $M$--$H$ curve at $T=4.5$\,K,
wherein one can note a large field interval of reversible magnetization
beyond $H_{irr}$. It could be argued that the linear extrapolation
of the nearly reversible magnetization beyond $H_{irr}$ to the $M=0$
axis would lead to an estimate of the upper critical field $H_{c2}$,
in accordance with the Ginzburg-Landau theory \cite{key-55}. However,
\begin{figure}[!t]
\begin{centering}
\includegraphics[scale=0.48]{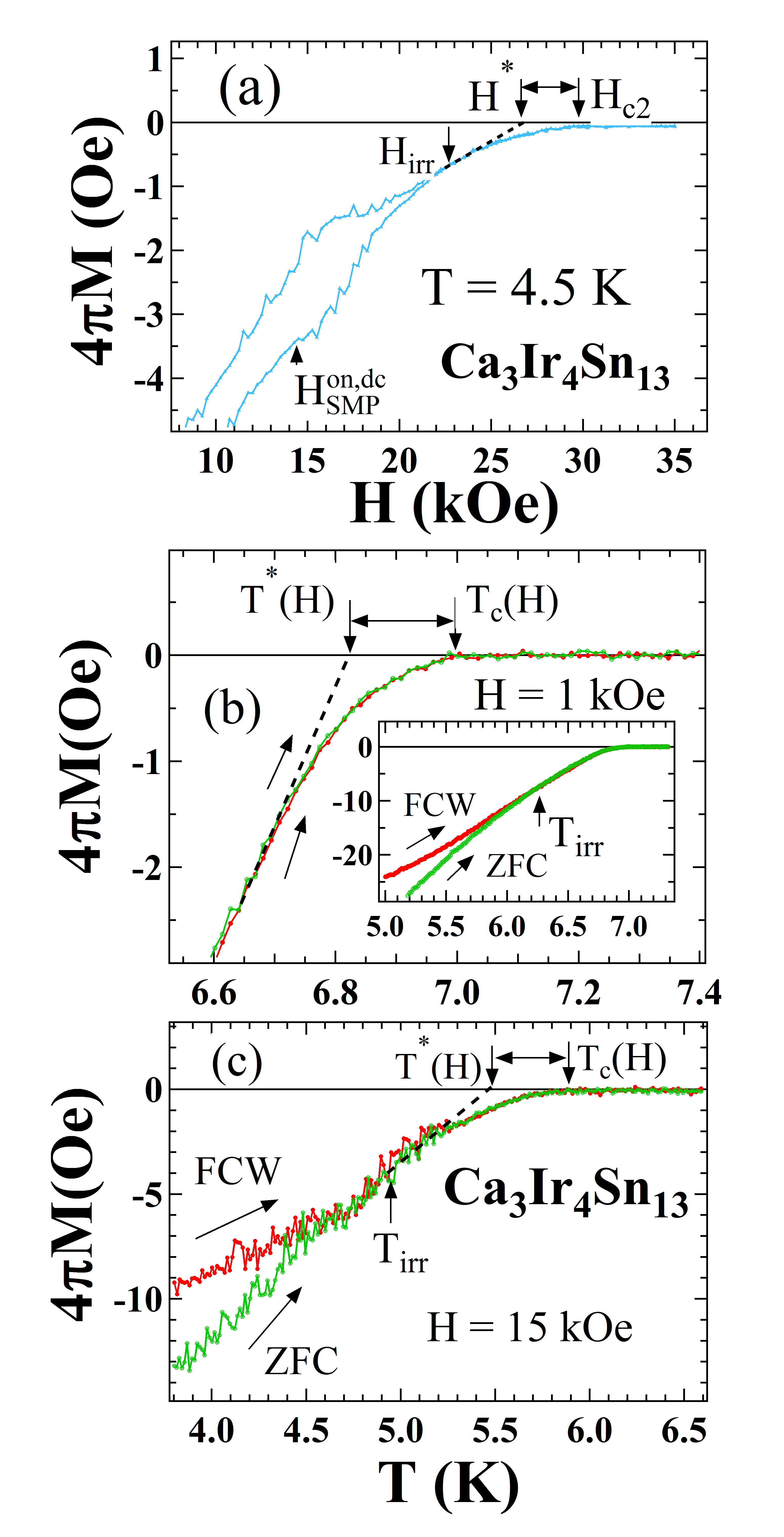}
\par\end{centering}

\caption{{\footnotesize (Color online) (a) Portion of $M$--$H$ curve at $T=4.5$\,K
near the superconducting-normal transition. The $M$--$H$ curve when
linearly extrapolated (dotted lines) beyond $H_{irr}$, yields a characteristic
field $H^{*}$. The onset of diamagnetism is marked as $H_{c2}$ ($>H^{*}$).
Portions of $ZFC$ and $FCW$ curves obtained at $H=1$\,kOe (panel
(b)) and $H=15$\,kOe (panel (c)). An inset in panel (b) shows the
merger of $ZFC$ and $FCW$ occurring at $T_{irr}$ for $H=1$\,kOe.}}
\end{figure}
in the present case, the linear extrapolation of magnetization data
beyond $H_{irr}$ to the $M=0$ axis identifies another characteristic
field, $H^{*}(T)$, which is significantly lower than the $H_{c2}$
value marked in Fig. 5(a).

To further comprehend the behavior in $M(H)$ response between $H_{irr}$
and $H_{c2}$, we now draw attention to the isofield dc magnetization
$M(T)$ data plotted in Figs.~5(b) and 5(c). These two panels show
the zero field-cooled ($ZFC$) and field-cooled warm-up ($FCW$) magnetization
responses at $H=1$\,kOe (Fig.~5(b)) and $H=15$\,kOe (Fig.~5(c)).
The sample was cooled in nominal zero field down to $T\sim2$\,K,
a field was applied and the magnetization was recorded while warming
to obtain $M_{ZFC}(T)$. The sample was then cooled from $T>T_{c}$
in the same field down to $T\sim2$\,K and the magnetization was
recorded while warming, which is the $M_{FCW}(T)$. It can be seen
that the $M_{ZFC}(T)$ and $M_{FCW}(T)$ curves merge at a temperature,
nominally marked as the irreversibility temperature $T_{irr}$ (see
the inset in Fig.~5(b), where this temperature is marked for $H=1$\,kOe).
Usually, a linear extrapolation of the (equilibrium) magnetization
above $T_{irr}$ to the $M=0$ axis is taken to identify the superconducting
transition temperature $T_{c}(H)$. However, similar to the $M(H)$
response near the superconducting transition (cf. Fig.~5(a)), there
exists a nearly reversible diamagnetic response persisting up to a
higher temperature, which we have marked as the critical temperature,
$T_{c}(H)$. Thus, the linear extrapolation of $M(T)$ curve above
$T_{irr}$ to the $M=0$ has been identified with a characteristic
temperature $T^{*}(H)$ ($<T_{c}(H)$) in Figs.~5(b) and 5(c). 
\begin{figure*}[!t]
\begin{centering}
\includegraphics[scale=0.5]{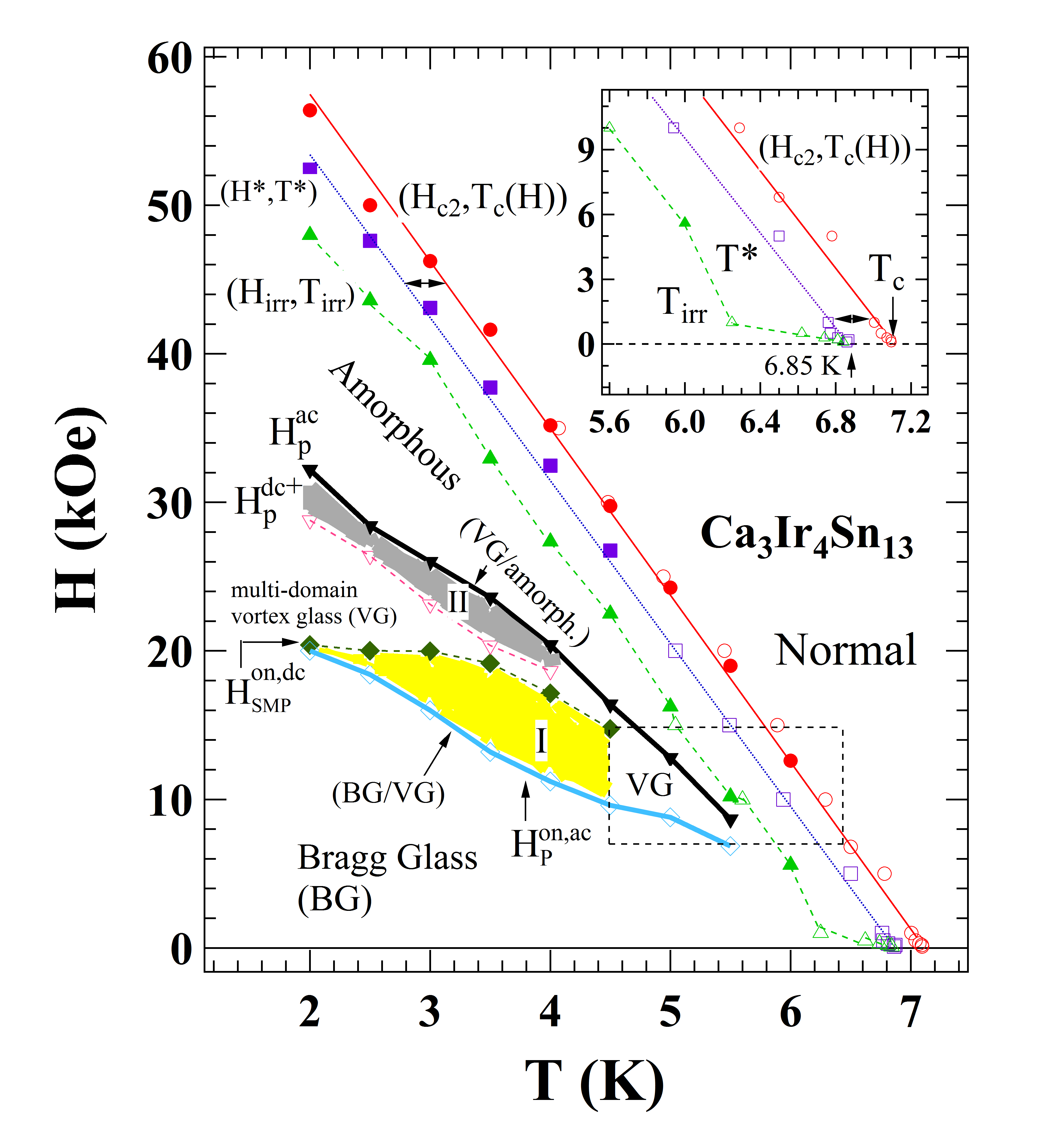}
\par\end{centering}

\caption{{\footnotesize (Color online) Vortex phase diagram in the given Ca$_{3}$Ir$_{4}$Sn$_{13}$
crystal, comprising phase boundaries obtained from both ac and dc
magnetization data. Region-I enclosed between the onset of SMP ($H_{SMP}^{on,dc}(T)$
extracted from dc $M$--$H$ scans) and the onset of PE ($H_{p}^{on,ac}(T)$
obtained from ac $\chi^{\prime}(H)$) corresponds to the ($H$--$T$)
phase space where an ac drive promotes disordering. Contrary to this,
in the region-II bounded by the peak fields ($H_{p}^{ac}(T)$ obtained
from ac $\chi^{\prime}(H)$) of the PE and the SMP ($H_{p}^{dc+}(T)$
extracted from dc $M$--$H$ scans) anomalies, an imposition of an
ac drive shows an improvement in ordering. The characteristics field
$H^{*}(T)$ (as obtained in Fig.~5) lies between $H_{irr}$ and $H_{c2}$
as shown. Inset shows that $H^{*}$ and $H_{c2}$ remain well separated
even in the close proximity of $T_{c}$..}}
\end{figure*}

\section{Discussion: The Vortex phase diagram}

The results of both ac and dc magnetization measurements can finally
be summarized by constructing a field-temperature ($H$--$T$) phase
diagram for a given crystal of Ca$_{3}$Ir$_{4}$Sn$_{13}$. Figure~6
illustrates this phase diagram depicting the various vortex phases
separated by the boundaries/crossover regimes, as indicated. The phase
boundary corresponding to the onset of the peak anomaly in $j_{c}(H)$
($H_{p}^{on,ac}$), as extracted from the ac $\chi^{\prime}(H)$ plots,
lies significantly below than the onset of the SMP ($H_{SMP}^{on,dc}$)
line obtained from the dc $M(H)$ curves. However, considering that
both the SMP and the PE anomalies signify order-disorder transformation(s)
in the vortex state, the region\,-\,I enclosed between the onset
field values of the two anomalies, (viz. between $H_{p}^{on,ac}(T)$
and $H_{SMP}^{on,dc}(T)$ lines in the main panel of Fig.~6), is
to be viewed as a well ordered vortex state via the dc $M(H)$ measurements.
The same field interval construes as a disordered region in view of
the anomalous response observed in this region in the ac $\chi^{\prime}(H)$
measurements. The superposed ac magnetic field in the $\chi^{\prime}(H,T)$
measurements per se acts as an additional driving force on the pinned
vortex matter. Such a driving force is typically known to transform
a metastable vortex configuration towards an equilibrium configuration
in a given ($H,T$) region \cite{key-26,key-56} by virtue of shaking
the vortices around their mean locations. It is pertinent to recall
here that from small angle neutron scattering measurements on the
vortex matter in a weakly pinned crystal of Nb, Ling $et$ $al.$
\cite{key-26}, had vividly shown that metastable disordered/ordered
states possible below/above the onset of the PE boundary transform
to their equilibrium ordered/disordered states in the respective ($H,T$)
regions on shaking with an ac drive. In this context, in the present
situation in Ca$_{3}$Ir$_{4}$Sn$_{13}$, the observation of disordered
configuration (region\,-\,I) in the ac $\chi^{\prime}(H)$ measurements
below the onset position of the SMP line is counter-intuitive. It
would be more appropriate to state that the ordered state is an equilibrium
state only below $H_{p}^{on,ac}(T)$ (shown by thick line in Fig.~6).
The ordered configuration observable in dc magnetization measurements
between $H_{p}^{on,ac}(T)$ and $H_{SMP}^{on,dc}(T)$ can thus be
identified with the notion of superheating of the BG phase above the
$H{}_{p}^{on,ac}(T)$ line. The said superheated BG state transforms
to its disordered version above the first order like $H{}_{p}^{on,ac}(T)$
line under the influence of an ac driving force in region\,-\,I.

In Fig.~6, it is instructive to note further that $H_{p}^{ac}(T)$
line lies above the $H_{p}^{dc+}(T)$ line (the latter obtained from
forward legs of the $M(H)$ curves, cf. Fig. 4(b)). If the ($H,T$)
region above $H_{p}^{dc+}(T)$ line is to be viewed as an amorphous
vortex matter, the $H$--$T$ phase space (i.e., region II) bounded
between $H_{p}^{dc+}(T)$ and $H_{p}^{ac}(T)$ lines is such that
an ac driving force is seen to improve the state of spatial order
in the vortex matter in this region. This is in sharp contrast to
the situation in region\,-\,I. On the basis of effect of an ac drive
in this region (II), one can also argue that the peak field of the
SMP ($H_{p}^{dc+}$) marks the onset of another order-disorder transition;
such an inference is possible only by combining the outcomes of the
influence of an ac drive with those obtained from $M(H)$ data, where
such a drive is not present. The amorphised state of the vortex matter
as an equilibrium state thus emerges above the $H_{p}^{ac}(T)$ values
(thick line in Fig.~6), instead of the usually presumed $H_{p}^{dc+}(T)$
values (dotted line in Fig. 6), as marked from the dc $M$--$H$ scans.
It may be added here that the $H{}_{p}^{on,ac}(T)$ and $H{}_{p}^{ac}(T)$
lines are independent of frequency in the range $10$\,Hz to $10^{3}$\,Hz
in which we repeated the ac susceptibility measurements in our set
up.

Considering that the onset ($H_{SMP}^{on,dc}(T)$) of SMP is nearly
temperature independent (for $T<4.5$\,K) and is located far below
the $H_{c2}$ line in Fig. 6, we are lead to surmise that the $H$--$T$
region between $H_{SMP}^{on,dc}$ and $H_{p}^{dc+}$ as the multi-domain
vortex glass (VG) state \cite{key-4,key-5,key-20,key-44}, in comparison
to the assigned nomenclature of Bragg glass (BG) to the vortex matter
below the onset position of the anomalous variation in critical current
density. In the absence of an ac drive, the BG extends up to $H_{SMP}^{on,dc}(T)$
line, whereas in the presence of a drive, BG state would get restricted
up to $H_{p}^{on,ac}(T)$ line. We noted above that for $T<4.5$\,K,
the $H$--$T$ phase space between $H_{p}^{on,ac}(T)$ and $H_{p}^{ac}(T)$
lines can be subdivided notionally into three parts on the basis of
effects of an ac drive on the underlying state of order in the vortex
matter. Above $T=4.5$\,K, we did not observe the fingerprint of
SMP anomaly in the dc $M$--$H$ data. The ($H$--$T$) region in
dotted rectangular box in Fig.~6 pertains to ($H$--$T$) phase space,
where we observe only the PE anomaly in the ac $\chi\prime$ data.
In the boxed region, $H{}_{p}^{on,ac}(T)$ line continues to mark
the BG to multi-domain VG transition and $H{}_{p}^{ac}(T)$ marks
the amorphization of the vortex solid within the domains in response
to the collapse of its elasticity. In this region ($T>4.5$\,K and
$6.8$\,kOe$<H<15$\,kOe), the $H{}_{p}^{ac}(T)$ line is in closer
proximity of the irreversibility ($H_{irr}$,$T_{irr}$) line. We
recall that in a weakly-pinned conventional superconductors, where
the PE phenomenon extends over a narrow ($H$,$T$) region and the
irreversibility line lies in close proximity of the peak field/temperature
of the PE, peak field of PE and/or the irreversibility line can be
identified as the melting line. The vortex matter in between $H{}_{p}^{ac}(T)$
and $H_{irr}(T)$ is often termed to be pinned liquid. Therefore,
in the boxed region ($T>4.5$\,K and $6.8$\,kOe$<H<15$\,kOe),
the $H_{irr}(T)$ could be accepted as the melting line. In ($H$,$T$)
domain, where the peak field/temperature of the PE is located far
away from the irreversibility line (as in the interval $T<4.5$\,K
and $H>15$\,kOe), the vortex matter above the peak field of the
SMP/PE perhaps comprises an admixture of pinned liquid and amorphous
solid phases.
\begin{table*}[t]
\caption{Superconducting parameters estimated for Ca$_{3}$Ir$_{4}$Sn$_{13}$
using the GL theory at $T=2$\,K}

\centering{}{\small }%
\begin{tabular}{|c|c|c|c|c|c|c|}
\hline 
{\small $\lambda$} & {\small $\xi$} & {\small $H_{c1}$} & {\small $H_{c2}$} & {\small $\kappa$} & {\small $G_{i}$} & {\small $j_{c}/j_{0}$ (at 15\,kOe)}\tabularnewline
\hline 
\hline 
{\small 2712\,$\textrm{\AA}$} & {\small 76\,$\textrm{\AA}$} & {\small 80 Oe} & {\small 56.4 kOe} & {\small $\sim35$} & {\small $\sim1.5\times10^{-8}$} & {\small $\sim10^{-5}$}\tabularnewline
\hline 
\end{tabular}
\end{table*}

We now focus attention on to the reversible region bounded in between
($H_{irr},T_{irr}$) and ($H_{c2},T_{c}$) lines in the main panel
of Fig.~6. It can be seen that the $H_{c2}$ enhances in a linear
way as temperature decreases below the superconducting transition
temperature of $7.1$\,K. The irreversibility line also displays
linear variation below $T\sim6.2$\,K. The inset in Fig.~6 shows
that above $T=6.2$\,K, the irreversibility line has a long tail,
which appears to terminate near $T=6.85$\,K as $H\rightarrow0$.
In Fig.~6, we have also chosen to plot the ($H^{*},T^{*}$) data
points obtained from linear extrapolation of the reversible magnetization
response, they lie in between the irreversibility and the $H_{c2}(T)$
lines. It is instructive to note that $H^{*}(T)$ runs parallel to
the $H_{c2}(T)$ line and it also meet the $T=0$ axis at $T\sim6.85$\,K,
distinct from the $T_{c}$ value of $7.1$\,K. If one asserts that
the reversible magnetization represents an equilibrium magnetization
response and concerns with the mean field description of a type\,-\,II
superconductor which predicts linear magnetization response, one may
conclude that the superconducting\,-\,normal transition ought to
terminate at $H^{*}/T^{*}$, rather than its extension up to a higher
field ($H_{c2}$)/higher temperature ($T_{c}$) as apparent from the
$M(H)/M(T)$ plots (cf. Fig.~5). The deviation from the mean field
description near the superconducting transition in the case of Ca$_{3}$Ir$_{4}$Sn$_{13}$
deserves a more detailed understanding. The recent observation of
ferromagnetic spin fluctuations in Ca$_{3}$Ir$_{4}$Sn$_{13}$ \cite{key-47}
could be advanced as a rationale for these deviations, however, such
a surmise may get discounted from the argument that the application
of high fields ought to suppress the spin fluctuations, whereas in
the present case, the separation between $H^{*}(T)$ and $H_{c2}(T)$
lines actually enhances at larger field values as apparent from Fig.~6.

\section{Summary}

To summarize, we have explored the vortex phase diagram in a weakly-pinned
single crystal of a low $T_{c}$ superconductor, Ca$_{3}$Ir$_{4}$Sn$_{13}$,
which is attracting adequate attention in the recent literature for
its novel physics. Table-1 contains various superconducting parameters
evaluated (at $T=2$\,K) for Ca$_{3}$Ir$_{4}$Sn$_{13}$, using
the Ginzburg-Landau theory. The weak pinning nature of this compound
is reflected by the ratio of depinning and depairing current densities
($j_{c}/j_{0}$) \textasciitilde{} 10$^{-5}$. The smallness of this
ratio may be attributed to the presence of point defects and metallurgical
source \cite{key-57} of disorder in such compounds. The ac susceptibility
($\chi^{\prime}(H,T)$) measurements have revealed the occurrence
of PE phenomenon in this compound. On the other hand, the results
of dc magnetization measurements reflect features typically identified
with the second magnetization peak anomaly. The phase boundaries constructed
by both the ac as well as the dc data prompt us to identify the metastable
regions in the vortex phase diagram of this compound. The region lying
between the onset lines of the SMP ($H_{SMP}^{on,dc}(T)$) in dc data
and the PE like phenomenon ($H_{p}^{on,ac}(T)$) in the ac data is
taken as a superheated ordered vortex matter, which sees an enhancement
in spatial disordering under the influence of an ac field. In contrast
to this, the region lying between the peak field of the PE ($H_{p}^{ac}$)
and the SMP ($H_{p}^{dc+}$) anomalies exhibits enhancement in the
ordering due to the effect of an ac drive. Such an effect ceases at
higher temperatures ($T>4.5$\,K) because of the absence of the SMP
feature there, only PE stands depicted at $T>4.5$\,K and for field
values lying between $5$\,kOe\, and$15$\,kOe. The region of the
$H$--$T$ phase space over which the anomalous variation in $j_{c}(H)$
predominates, is very broad and it bears resemblance with anomalous
data in several other low $T_{c}$ as well as high $T_{c}$ superconductors,
including the Cuprates. The broad (H,T) region comprises per se two
transitions, a Bragg glass (BG) to vortex glass (VG) transition commencing
at $H_{p}^{on,ac}(T)$ (as in Fig.~6 for Ca$_{3}$Ir$_{4}$Sn$_{13}$)
and the amorphization of the vortex matter commencing at $H_{p}^{dc+}(T)$
(as in Fig.~6). The $H_{p}^{ac}(T)$ line for Ca$_{3}$Ir$_{4}$Sn$_{13}$
in Fig.~6 can be taken to mark the spinodal line of the second transition.
The gap between $H_{p}^{dc+}$ and $H{}_{p}^{ac}$ values progressively
decreases as T approaches $4.5$\,K, and at $5$\,K, the said difference
ceases to exist as the PE in the ac susceptibility subsumes the notion
of SMP. 

In the end, we may point out that the vortex phase diagram (Fig.~6)
of Ca$_{3}$Ir$_{4}$Sn$_{13}$ also raises an important issue regarding
the source of diamagnetic response between $H^{*}/T^{*}$ and $H_{c2}/T_{c}$.
As per the G\,-\,L theory, the expected upper critical field/temperature
should have been the ($H^{*},T^{*}$) line in Fig.~6, the experimental
values of the same are slightly enhanced to ($H_{c2},T_{c}$) line.
There is a significant separation between the two lines (($H^{*},T^{*}$)
and ($H_{c2},T_{c}$)) even in an infinitesimally small field, as
they intersect the temperature axis at $T\sim6.85$\,K and $T_{c}$,
respectively. It is also pertinent to note that the G\,-\,L theory
does not explicitly take into account the spatial ordering of the
vortex matter. We have witnessed a reversible (amorphous) region (in
the dc $M(H)/M(T)$ response) that emerged from the disordered configuration
(PE) of the vortex matter, wherein (i.e., above ($H^{*},T^{*}$) line)
the mean field description does not hold. A possible source for such
an unusual observation could be the existence of surface superconductivity
\cite{key-58} above the $H^{*}/T^{*}$ line and surviving up to the
$H_{c2}/T_{c}$ line in the $H$--$T$ phase space. Efforts to explore
this region of vortex phase diagram are currently in progress \cite{key-59},
with some significant success in sight. 

\noindent \textbf{\large Acknowledgments}{\large \par}

\noindent Santosh Kumar would like to thank the Council of Scientific
and Industrial Research, India for grant of the Senior Research Fellowship.

\textbf{\large References}{\large \par}

\end{document}